\documentclass{article}

\usepackage{arxiv}
\usepackage{natbib}
\setcitestyle{square,sort,comma,numbers}
\usepackage[utf8]{inputenc} 
\usepackage[T1]{fontenc}    
\usepackage{hyperref}       
\usepackage{url}            
\usepackage{booktabs}       
\usepackage{amsfonts}       
\usepackage{nicefrac}       
\usepackage{microtype}      
\usepackage{lipsum}
\usepackage{framed,multirow}
\usepackage{booktabs}
\usepackage{diagbox}
\usepackage{algorithm}
\usepackage{algorithmic}%
\usepackage{bbm}
\usepackage{tabularx}
\usepackage{lscape}
\usepackage{lineno}
\usepackage{amssymb}
\usepackage{latexsym}
\usepackage{amsmath}
\usepackage{subfigure}
\usepackage{booktabs}
\usepackage{placeins}
\usepackage{hyperref}
\usepackage{array}
\usepackage{float}
\usepackage{pdfcomment}
\usepackage{graphics}
\usepackage{graphicx}
\usepackage{xcolor}
\definecolor{newcolor}{rgb}{.8,.349,.1}
\definecolor{newcolor}{rgb}{.8,.349,.1}
\definecolor{mygreen}{rgb}{0.1, 0.75, 0.15}
\newcommand{\norme}[1]{\left\Vert #1\right\Vert}
\newcommand{\argmin}{\operatornamewithlimits{argmin}}

\newcommand{\tabitem}{~~\llap{\textbullet}~~}

\newcommand{\removeGG}[1]{}

\newcommand{\removeRH}[1]{}

\usepackage{tikz}
    \usetikzlibrary{shapes.arrows}
    \usetikzlibrary{decorations.shapes}
    \usetikzlibrary{decorations.pathreplacing}
    \usetikzlibrary{fadings,shapes.arrows,shadows}
\makeatletter
\DeclareRobustCommand\sfrac[1]{\@ifnextchar/{\@sfrac{#1}}%
                                            {\@sfrac{#1}/}}
\def\@sfrac#1/#2{\leavevmode\kern.1em\raise.5ex
         \hbox{$\m@th\mbox{\fontsize\sf@size\z@
                           \selectfont#1}$}\kern-.1em
         /\kern-.15em\lower.25ex
          \hbox{$\m@th\mbox{\fontsize\sf@size\z@
                            \selectfont#2}$}}

\usetikzlibrary{positioning}

\title{Data-driven audio recognition: a supervised dictionary approach}

\author{
  Imad Rida  \\
 Laboratoire BMBI Compiègne \\ 
 Université de Technologie de Compiègne \\ 
  Compiègne, France\\
  }

\begin{document}
\maketitle

\begin{abstract}
Machine hearing or listening represents an emerging area. Conventional approaches rely on the design of handcrafted features specialized to a specific audio task and that can hardly generalized to other audio fields. Unfortunately, these predefined features may be of variable discrimination power while extended to other tasks or even within the same task due to different nature of clips. Motivated by this need of a principled framework across domain applications for machine listening, we propose a generic and data-driven representation learning approach. For this sake, a novel and efficient supervised dictionary learning method is presented. Experiments are performed on both computational auditory scene (East Anglia and Rouen) and synthetic music chord recognition datasets. Obtained results show that our method is capable to reach state-of-the-art hand-crafted features for both applications
\end{abstract}


Humans have a very high perception capability through physical sensation, which can include sensory input from the eyes, ears, nose, tongue, or skin. A lot of efforts have been devoted to develop intelligent computer systems capable to interpret data in a similar manner to the way humans use their senses to relate to the world around them. While most efforts have focused on vision perception which represents the dominant sense in humans, machine hearing also known as machine listening or computer audition represents an emerging area \citep{lyon2010machine}.

Machine hearing represents the ability of a computer or machine to process audio data. There is a wide range variety of audio application domains including music, speech and environmental sounds. Depending on the application domain, several tasks can be performed such as, speech/speaker recognition, music transcription, computational scene auditory recognition, etc (see Table \ref{hearingapplication}).

\begin{table}[h]
\setlength\extrarowheight{13.8pt}
\centering
\caption{Machine hearing tasks based on different application domains } 
\label{hearingapplication}
\resizebox{\columnwidth}{!}{%
\begin{tabular} {lp{4cm}p{4cm}p{4cm}}
\diagbox{\textbf{Tasks}}{\textbf{Domains}}& \textbf{Environnemental} &\textbf{Speech} & \textbf{Music}     \\
\textbf{Description}  & {\footnotesize Environment}    & {\footnotesize Emotion} & {\footnotesize Music Recommendation }                                        \\
\textbf{Classification} &  {\footnotesize Computational Auditory Scene Recognition} & {\footnotesize Speech or Speaker Recognition}     & {\footnotesize Music Transcription}   \\
\textbf{Detection}    & {\footnotesize Event Detection }& {\footnotesize Voice Activity Detection}    &{\footnotesize Music Detection }   \\

\end{tabular}}

\end{table} 


In this chapter, we are interested in the classification of audio signals in both environmental and music domains and more particularly, Computational Auditory Scene Recognition (CASR) and music chord recognition. The former refers to the task of associating a semantic label to an audio stream that identifies the environment in which it has been produced while the second task aims to recognize music chords that represent the most fundamental structure and the back-bone of occidental music. 

In the following we briefly review different approaches for audio signal classification. A novel method able to learn the discriminative feature representations will be introduced. Extensive experiments will be performed on CASR and music chord benchmark databases and results will be compared to conventional state-of-the-art hand-crafted features.

\section{Problem statement}
The problem of audio signal classification is now becoming more and more frequent, ranging from 
speech to non-speech signal classification. The usual trend to classify signals is first to extract discriminative feature representations from the signals, and then feed a classifier with them. Features are chosen so as to enforce similarities within a class and disparities between classes. The more discriminative the features are, the better the classifier performs. 

For each audio signal classification problem, specific hand-crafted features have been proposed. For instance, chroma vectors represent the dominant representation which has been developed in order to extract the harmonic content from music signals for different applications \citep{oudretemplate,oudre2011chord,chroma,sheh2003chord,mauch2010simultaneous,ellis2007classifying,miotto2008music,bartsch2005audio}.

In audio scene recognition, recorded signals can be potentially composed of a very large amount of  sound events while only few of these events are informative. Furthermore, the sound events can be from different nature depending on the location (street, office, restaurant, train station, etc). To tackle this problem, features such as Mel-Frequency Cepstral Coefficients (MFCCs)  \citep{davis1980comparison, kinnunen2012low,zheng2001comparison,benzeghiba2007automatic,li2014overview,rida2018feature} have been successfully applied and combined with different classification techniques \citep{ellis2004computational, peltonen2002computational}.
  
These predefined features may be of variable discrimination power according to the signal nature and learning task if they are extended to other application domains. For this reason machine hearing systems should be able to learn automatically the suited feature representations. Time-frequency features have shown good ability to represent real-world signals \citep{davy2002optimized} and methods have been designed to learn them. They can be broadly divided into four main approaches \citep{sangnier2015filter}: wavelets, Cohen distribution design, dictionary and filter banks learning summarized in Table \ref{tab:summtff}.

\begin{table}[h]
  \centering
  \caption {Non exhaustive time-frequency representation learning for classification  \citep{sangnier2015filter}.}
  \begin{tabular}{llll}
    \multicolumn{2}{c}{}  \\
        \toprule
         \toprule
    Approach & Methods   \\
    \midrule  
     \midrule  
                                   & \tabitem \citep{jones2001genetic}  \\
     \tabitem  Wavelets & \tabitem \citep{strauss2003feature} \\
                                   &  \tabitem \citep{yger2011wavelet} \\
                                   \hline
                                   \hline
    \tabitem Cohen Distribution  & \tabitem \citep{davy2002optimized}  \\
                                                  &  \tabitem \citep{honeine2006optimal} \\
                                     \hline
                                     \hline
     \tabitem Dictionary   & \tabitem \citep{mairal2009}  \\
                                      & \tabitem \citep{ramirez2010}  \\
                                      \hline
                                      \hline
     \tabitem Filter Bank     &  \tabitem \citep{biem2001application}  \\
                                         &  \tabitem \citep{sangnier2015filter}  \\
                                         
    \bottomrule
     \bottomrule
  \end{tabular}
  \label{tab:summtff}
\end{table}

Wavelets showed very good performance in the context of compression \citep{tewfik1992optimal,claypoole1998adaptive} where one minimizes the error between the original and approximate signal representation. While the latter may be a salutary goal, it does not well address the classification problems. \citep{jones2001genetic} suggested a classification-based cost function maximizing the minimum probability of correct classification along the confusion-matrix diagonal. This cost function is optimized using a genetic algorithm (GA) \citep{goldberg1988genetic}. \citep{strauss2003feature} tried to tune their introduced wavelet by maximizing the distance in the wavelet feature space of the means of the classes to be classified. This is done by constructing a shape-adapted Local Discriminant Bases (LDBs) called also morphological LDBs (MLDBs) as an extension of LDBs \citep{saito2002discriminant}. In other words they aim to select bases from a dictionary that maximize the dissimilarities among classes. \citep{yger2011wavelet} tried to learn the shape of the mother wavelet, since classical wavelet such as Haar, or Daubechies ones may not be optimal for a given discrimination problem. Then, the best wavelet coefficients that are useful for the discrimination problem are selected. Features obtained from different wavelet shapes and coefficient selections were combined to learn a large-margin classifier.

In the Cohen distribution design, \citep{davy2002optimized}  proposed to use a Support Vector Machine (SVM) of the Cohen's group Time-Frequency Representations (TFRs). The main problem is that the classification performance is depending on the choice of TFR and SVM kernel respectively. To tackle this problem, they presented a simple optimization procedure to determine the optimal SVM and TFR kernel parameters. \citep{honeine2006optimal} proposed a method for selecting Cohen class time-frequency distribution appropriate for classification tasks based on the kernel-target alignment \citep{cristitiaini2002kernel}. 

Motivated by their success in image denoising \citep{elad2006,rida2016gait} and inpainting \citep{elad2010}, dictionary learning was further extended to classification tasks. It consists in finding a linear decomposition of a raw signal or potentially its time-frequency representation using a few atoms of a learned dictionary. While conventional dictionary learning techniques tried to minimize the signal reconstruction error, \citep{mairal2009,mairal2012task} introduced supervised dictionary by embedding a logistic loss function to simultaneously learn a classifier, the dictionary $\mathbf{D}$ and the decomposition coefficients of the signals over $\mathbf{D}$. \citep{ramirez2010} introduced a dictionary learning method by adding a structured incoherence penalty term to learn $C$ dictionaries for $C$ classes while enforcing incoherence in order to make these dictionaries as different as possible.

In the filter bank approach, \citep{biem2001application} designed a method named Discriminative Feature Extraction (DFE) where both the feature extractor and classifier are learned with the objective to minimize the recognition error. The designed feature extractor is a filter bank where each
filter's frequency response has a Gaussian form determined by three kinds of parameters (center frequency, bandwidth, and gain factor). The classifier was defined as a prototype-based distance \citep{mcdermott1994prototype}. \citep{sangnier2015filter} proposed to build features by designing a data-driven filter bank and by pooling the time-frequency representations to provide time-invariant features. For this purpose, they tackled the problem by jointly learning the filters of the filter bank with a support vector machine. The resulting optimization problem boils down to a generalized version of a Multiple Kernel Learning (MKL) problem \citep{rakotomamonjy2008simplemkl}.

\bigskip

It can be seen that methods among, wavelets, Cohen distribution and filter bank approaches, solely seek to find a suitable time-frequency feature representation for signal classification. Although time-frequency representations showed efficiency to classify temporal signal (audio, electroencephalography, etc), there is no effectiveness guarantee for all type of signals. On the other side, dictionary learning can be combined with any initial feature representation and hence may have the ability and flexibility to deal with signals from different nature. Indeed, supervised dictionary learning can be seen as task-driven approach for learning discriminative representations.

In this chapter, based on an initial time-frequency representation, the problem of signal audio recognition is formulated as a supervised dictionary learning problem. The resulting optimization problem is non-convex and solved using a proximal gradient descent method. In the following we introduce our representation learning method based on dictionary learning as well as the performed experiments on both music chord recognition and computation auditory scene recognition databases. 

\section{Dictionary learning for audio signal classification}
Sparse representation of signals and images has known a big interest from researchers in order to analyze, extract or select features.  A "sparse representation" means that a signal or image can be represented as a linear combination of few representative elements, called dictionary atoms. The main  challenge of the sparse representation is the choice of the dictionary on which the signal will be represented and the sparsity type. The simplest approach to tackle this problem is to take predefined dictionary such as wavelet analysis, Gabor atoms or Discrete Cosine Basis, but this will give us no guarantee that these predefined dictionaries will be able to represent and extract useful information for the problem in question. 

Alternative approach is to learn the suited set of atoms from the data. From the view of compression sensing, dictionary learning is originally designed to learn an adaptive codebook to faithfully represent the signals with sparsity constraint. Dictionary learning has been applied for different applications such as image denoising \citep{elad2006,mairal2008}, inpainting \citep{elad2010,mairal2008}, clustering \citep{cheng2010,wright2010} and classification \citep{bradley2008,mairal2009,mairal2012task,rida2018palmprint,rida2015human,al2019palmprint,rida2018ensemble,rida2019palmprint,rida2016robust,rida2015unsupervised}. 

In the following we review the conventional dictionary learning based on a single dictionary and the different approaches to build supervised dictionary for classification. We also introduce our class based dictionary learning method.

\subsection {Conventional dictionary learning}

Let suppose a dictionary $\mathbf{D} \in \mathbb {R}^{M \times K}$ composed of $K$ atoms $\{\mathbf{d}_{k} \in \mathbb{R}^{M}\}_{k=1}^{K} $. We seek a sparse representation $\mathbf{a}_{n} \in \mathbb{R}^{K}$ of a signal $\mathbf{x}_{n} \in \mathbb {R}^{M}$ over $\mathbf{D}$ such as:\\

\begin{equation}
\mathbf{x}_{n} \approx \sum_{k=1}^{K} a_{nk} \mathbf{d}_{k} = \mathbf{D} \mathbf{a}_{n}
 \end{equation}

Given a set of $N$ signals $\{\mathbf{x}_{n}\}_{n=1}^{N}$, the coefficients of $\mathbf{a}_{n}$ as well as the dictionary $\mathbf {D}$ are obtained by solving the following optimization problem:\\
 
 \begin{equation}
 \left\{
    \begin{array}{ll}
       \displaystyle \min_{\mathbf{D}, \{\mathbf{a}_{n}\}_{n=1}^{N}} \sum_{n=1}^{N} \norme{\mathbf{x}_{n}-\mathbf{D} \mathbf{a}_{n}}_{2}^{2} + \lambda \norme{\mathbf{a}_{n}}_{1}\\
       & \\
   \displaystyle  \hspace{0.4cm} \mbox{s.t} \hspace {0.5 cm} \norme{\mathbf{d}_{k}}_{2}^{2} \leq 1  \hspace {0.5 cm} \forall k=1, \cdots, K 
   \end{array}
\right.
\label{eq:conventional}
\end{equation}

It can be seen that the original formulation for dictionary learning is based on the minimization of the reconstruction error between a signal and its sparse representation over the learned dictionary.  Although this formulation is optimal for solving problems such as denoising and inpainting, it may not lead to optimal solution in classification tasks, where the ultimate goal is to make the learned dictionary and corresponding sparse representation as discriminative as possible since it does not take the label information in consideration. This motivated the emergence of supervised dictionary learning techniques \citep{rida2018comprehensive,rida2018novel,dehais2019pbci}.

\subsection {Supervised dictionary learning}
Supervised dictionary learning can be organized in six main groups \citep{gangeh2015supervised}: learning one dictionary per class, unsupervised dictionary learning followed by supervised pruning, joint dictionary and classifier learning, embedding class labels into the learning of dictionary, embedding class labels into the learning of sparse coefficients and learning a histogram of dictionary elements over signal constituents. In the following we briefly introduce these approaches as well the main works belonging to them. Note that the advantages and drawbacks of each approach are summarized in Table \ref{tab:summtf}.

\subsubsection{Learning one dictionary per class}

The first and simplest approach is to compute one dictionary per class, i.e., using the training samples of each class, a dictionary is constructed. The overall dictionary is obtained by the concatenation of individual class dictionaries. In this framework, \citep{wright2009robust} proposed the so-called Sparse Representation-based Classification (SRC), where training samples of each class serve as dictionary. The sparse representation of a testing sample over each dictionary is calculated based on Lasso. The test sample is then assigned to class label which dictionary provides the minimal residual reconstruction error. \citep{yang2010}, instead to use dictionaries based on training samples proposed to learn a dictionary per class based on the conventional approach (\ref{eq:conventional}). Although this approach can be potentially performing, learned dictionaries can capture similar properties for different classes leading to poor classification performance. To tackle this problem, \citep {ramirez2010} suggested to make the learned dictionaries as different as possible to capture distinct information by minimizing the pairwise similarity between dictionaries. \citep{kong2012dictionary} proposed to learn a dictionary per class to capture the particularity information and a shared dictionary to capture the commonality. After finding the overall dictionary, the classification of test samples is performed the same way as with the SRC.

\subsubsection{Prune large dictionaries}

In this approach, a very large dictionary in learned following the conventional approach  (\ref{eq:conventional}), then the dictionary atoms are merged based on a predefined criterion so as to obtain a reduced discriminative dictionary. For instance, \citep{fulkerson2008localizing} used Agglomerative Information Bottleneck (AIB) which iteratively merges two atoms that cause the smallest decrease in the mutual information between the dictionary atoms and the class labels. In the same context, \citep{winn2005object}  proposed another method based on merging two dictionary atoms so as to minimize the loss of mutual information between the histogram of dictionary atoms and class labels.

\subsubsection{Joint dictionary and classifier learning}
This approach showed very good performances and represented a big advance in the field. It seeks to jointly learn dictionary and classifier. In \citep{mairal2009} a linear classifier and logistic loss function was applied. \citep{zhang2010} suggested a technique called discriminative K-SVD (DK-SVD) which also jointly learns the classifier parameters and dictionary. However, instead to solve the optimization problem iteratively and alternately between classifier parameters and dictionary, a sub-optimal learning process is built upon two main steps. The first one aims to learn a conventional dictionary and sparse representation coefficients of the signals over it. The second step uses the resulting sparse coefficients to learn a linear classifier. 

\subsubsection{Embedding class labels into the learning of dictionary}

In this framework we can cite the approach of \citep{zhang2013simultaneous}. They propose to first project the data into an orthogonal space where the intra and inter-class reconstruction errors are minimized and maximized respectively, and subsequently learn the dictionary and the sparse representation of the data in this new space. \citep{lazebnik2009supervised} seek to minimize the information loss due to class labels prediction from a supervised learned dictionary instead of the original training data samples.

\subsubsection{Embedding class labels into the learning of sparse coefficients}
This approach seeks to include class labels in the learning of coefficients. Supervised coefficient is based on minimizing the within-class covariance of coefficients and at the same time maximizing their between-class covariance. \citep{yang2011} tried to learn simultaneously a dictionary per class by decomposing every signal $\mathbf{x}_{n}$ with label $y_{n}$ over the $C$ dictionaries and enforcing the sparsity of the coefficients related to the dictionaries $\mathbf{D}_{j}$ such that $y_{n} \neq j$. Classification of a new sample is done in the same way as SRC  \citep{wright2009robust}.

\subsubsection{Learning a histogram of dictionary elements over signal constituents}
There are situations where a signal is made of some local constituents, e.g., an image is made up of patches or a speech made of phonemes. In this case histogram of dictionary atoms learned on local constituents is computed. The resulting histograms are used to train a classifier and predict the class label of unknown

\begin{landscape}
\begin{table}[h]
  \centering
  \caption {Summary of supervised dictionary learning techniques for data classification \citep{gangeh2015supervised}.}
  \begin{tabular}{llll}
    \multicolumn{3}{c}{}  \\
        \toprule
         \toprule
    Methods & Approach  & Advantages  \& Drawbacks  \\
    \midrule  
     \midrule

    \citep{wright2009robust} &                                                &    \\
  \citep{yang2010}               &  \textbf{A.}~  Dictionary per class   &  $(+)$ ease dictionary computation   \\
    \citep {ramirez2010}        &                                                      &    $(-)$     very large dictionary \\
     \citep{kong2012dictionary}        &                                                      &    \\

                                   \hline
                                   \hline
                                   
 \citep{fulkerson2008localizing} &  \textbf{B.}~   Prune large dictionaries     & $(+)$ ease dictionary computation \\
\citep{winn2005object}                &                                                               & $(-)$ low performances  \\
                                     \hline
                                     \hline
                                     
\citep{mairal2009}     &   \textbf{C.}~  Joint dictionary \& classifier learning  &  $(+)$ good performances \\
\citep{zhang2010}     &                                                                                  & $(-)$ too many parameters     \\
                                     \hline
                                      \hline
                                      
  \citep{zhang2013simultaneous}  &   \textbf{D.}~   Labels in dictionary   & $(+)$ good performances \\
  \citep{lazebnik2009supervised}   &                                                         &  $(-)$ complex optimization  \\

                                         \hline
                                         \hline
                                         
 \citep{yang2011}     &   \textbf{E.}~     Labels in coefficients  &  $(+)$ good performances     \\
                                  &                              & $(-)$ complex \\
 
                                         \hline
                                         \hline

 \citep {varma2009statistical}                                                                               &    \textbf{F.}~  Histograms of dictionary elements   & $(+)$ good performances \\                                                                                                                                                                                                                                         
  \citep{lian2010probabilistic}           &   & $(-)$ only based local constituents   \\
                                                                                                                                                          
    \bottomrule
     \bottomrule
  \end{tabular}
  \label{tab:summtf}
\end{table}
\end{landscape}
\noindent signals. \citep {varma2009statistical} aggregated small patches over all images in a class, and clustered them using k-means algorithm. Obtained cluster centers form a dictionary. Although the latter method gives good results, it does not really include the label information in the learning process. This motivated to exploit the class information to learn dictionaries in supervised way \citep{lian2010probabilistic}. 

\bigskip

Based on the brief study of supervised dictionary approaches, we introduce in the following a novel supervised dictionary method. Our proposed method tries to exploit the strong points of the previous methods that is: i) learning one dictionary per class, and ii) embedding class labels to force sparse coefficients. To this end, we encourage the dissimilarity between the dictionaries by penalizing the pairwise similarity between them. To reach superior discrimination power, we push towards zero the coefficients of a signal representation over other dictionaries than the one corresponding to its class label. 

\subsection {Class based dictionary learning}
\label{sec:ourcontribution}
Let consider $\{(\mathbf{x}_{n},y_{n})\}_{n=1}^{N}$ where $\mathbf{x}_{n} \in \mathbb{R}^{M}$ is a signal and $y_{n} \in \{1,\cdots, C\}$ its label. We consider a dictionary $\mathbf {D}_{c} \in \mathbb {R}^{M \times K'}$ associated to each class $c$. The global dictionary $\mathbf{D}= [ \mathbf{D}_{1} \cdots \mathbf{D}_{C} ] \in \mathbb{R}^{M \times K}$ represents the concatenation of the class based dictionaries $\{\mathbf{D}_{c}\}_{c=1}^{C}$. Each dictionary $\mathbf{D}_{c}$ is composed of $K'$ atoms $\{\mathbf{d}_{k} \in \mathbb{R}^{M}\}_{k=1}^{K'} $. For simplicity sake we consider $K'$ is the same for all $\{\mathbf{D}_{c}\}_{c=1}^{C}$. The sparse representation of $\mathbf{x}_{n}$ over the global dictionary $\mathbf{D}$ is $\mathbf{a}_{n}^{T}= [ \mathbf{a}_{n1}^{T} \cdots \mathbf{a}_{nc}^{T} \cdots \mathbf{a}_{nC}^{T}]$ where $\mathbf{a}_{nc}$ represents the sparse representation over the class specific dictionary $\mathbf{D}_{c}$. Hence the sparse representation of the overall training data$\{\mathbf{x}_{n}\}_{n=1}^{N}$ is gathered in $\mathbf{A}=[ \mathbf{a}_{1} \cdots \mathbf{a}_{n}]$. The dictionary learning problem we intend to address is formulated as follows:

\begin{equation}
 \left\{
    \begin{array}{ll}
      \displaystyle  \min_{\{\mathbf{D}_{c}\}_{c=1}^{C}, \{\mathbf{a}_{n}\}_{n=1}^{N}} J = J_{1} + J_{2} + \lambda J_{3} +\gamma_{1} J_{4}+ \gamma_{2} J_{5}\\
       & \\
   \displaystyle  s.t \hspace {0.5 cm} \norme{\mathbf{d}_{ck}}_{2}^{2} \leq 1  \hspace {0.5 cm} \forall c=1, \cdots, C \hspace {0.5 cm} \mbox{and} \hspace {0.5 cm} \forall k=1, \cdots, K \\ \\
   \end{array}
\right.
\label{eq:2}
\end{equation}
where in the problem (\ref{eq:2})

 $$J_{1}=\displaystyle \sum_{n=1}^{N} \norme{\mathbf{x}_{n}-\mathbf{D} \mathbf{a}_{n}}_{2}^{2}$$
 \noindent represents the global reconstruction error over the global dictionary $\mathbf{D}$.
 
$$J_{2}=\displaystyle \sum_{c=1}^{C} \sum_{n=1}^{N} \mathbbm{1}_{y_{n}=c} \norme{\mathbf{x}_{n}-\mathbf{D}_{c} \mathbf{a}_{nc}}_{2}^{2}$$
 \noindent stands for the class specific reconstruction error over the dictionary $\mathbf{D}_{c}$. In other words $J_{2}$ measures the quality of reconstructing a sample $(\mathbf{x}_{n}, \mathbf{y}_{n}=c)$ over the sole dictionary $\mathbf{D}_{c}$.

$$J_{3}=\displaystyle  \sum_{n=1}^{N}  \norme{\mathbf{a}_{n}}_{1}$$
 \noindent  is the classical sparsity penalization.
 
$$J_{4}=\displaystyle \sum_{n=1}^{N} \sum_{c=1}^{C}  \mathbbm{1}_{y_{n} \neq c} \norme{\mathbf {a}_{nc}}_{2}^{2}$$
\noindent  aims to push toward zero the coefficients  $\mathbf{a}_{nc}$ of the signal $\mathbf{x}_{n}$ representation over non-class specific dictionary $\mathbf{D}_{j}$, $j \neq y_{n}$.

$$J_{5}=\displaystyle \sum_{c=1}^{C} \sum_{\substack {c'=1 \\ c' \neq c}}^{C} \norme {\mathbf{D}_{c}^{T} \mathbf{D}_{c'}}_{F}^{2}$$
 \noindent with $\norme{.}_{F}$ is the Frobenius norm, encourages the pairwise orthogonality between different dictionaries.
 
 \bigskip
 
To sum up, our dictionary learning problem (\ref{eq:2}) seek to: 

\begin{itemize}

\item Capture as much as possible information in the signal by minimizing the global reconstruction error. 
\item Specialize the extracted information per class by minimizing the class specific reconstruction error similar to intra-class variations minimization. 
\item Render dissimilar the extracted class specific information by promoting orthogonality of dictionaries and "zeroing" coefficients not specific to the sample label. In other words, we attempt to maximize inter-class variations.
\item Promote coefficients sparsity to maintain generalization ability.
\end{itemize}

$\lambda$, $\gamma_{1}$ and $\gamma_{2}$ are regularization parameters controlling the sparsity, the structure of sparse coefficients and pairwise orthogonality of learned dictionaries respectively. We could have associated a regularization parameter to the term $J_{2}$, however to avoid multiplying the number of hyper-parameters we choose to fix it to $1$. Furthermore, conducted experiments show that it does not have significant impact on the performances. 

\bigskip

Compared to \citep{kong2012dictionary} where they propose to learn a shared dictionary combined with class specific, we only rely on the latter one. Furthermore their optimization scheme is based on a simplifying assumption that $\mathbbm{1}_{y_{n} \neq c} \norme{\mathbf {a}_{nc}}_{2}^{2}=0$ which eases the optimization but harms the convergence. In our formulation we do not rely on those assumptions and we provide a more general optimization algorithm described in the next section.

\subsection{Optimization scheme}
\label{sec:opt}
At the first sight, the objective function in (\ref{eq:2}) seems to be complex but it can be solved based on an alternating optimization scheme which involves a sparse coding step and dictionary optimization step. Indeed, problem (\ref{eq:2}) is convex in $\mathbf{D}_{c}$ for the coefficients $\mathbf{a}_{nc}$ fixed and is so the inverse way when the $\mathbf{D}_{c}$ are fixed.

\subsubsection{Sparse coding step}
In this step, we fix $\{\mathbf {D}_{c}\}_{c=1}^{C}$ and we estimate the coefficients $\{\mathbf{a}_{n}\}_{n=1}^{N}$. For each signal $\mathbf {x}_{n}$ of class $y_{n}$, the related vector $\mathbf{a}_{n}$ is decoupled in the optimization problem.  Let $y_{n}=c'$, this conducts us to solve the following problem: \\
\begin{equation}
\displaystyle  \min_{\mathbf{a}_{n}}  \norme{\mathbf{x}_{n}-\mathbf{D} \mathbf{a}_{n}}_{2}^{2} +  \norme{\mathbf{x}_{n}-\mathbf{D}_{c'} \mathbf{a}_{nc'}}_{2}^{2} + \gamma_{1} (\norme {\mathbf{a}_{n}}^{2}_{2} - \norme {\mathbf{a}_{nc'}}^{2}_{2}) + \lambda \norme {\mathbf{a}_{n}}_{1}
\label{eq:3}
 \end{equation}
where $ \displaystyle \norme{\mathbf{a}_{n}}_{2}^{2}= \sum_{c=1}^{C} \norme{\mathbf{a}_{nc}}_{2}^{2}$ and $\displaystyle \sum_{c=1}^{C} \mathbbm{1}_{c \neq c'} \norme {\mathbf{a}_{nc}}_{2}^{2}= \norme{\mathbf{a}_{n}}_{2}^{2}- \norme{\mathbf{a}_{nc'}}_{2}^{2}$ \\

It can be seen that (\ref {eq:3}) consists of quadratic error terms and elastic-net type penalization. Thus this problem is amenable to a Lasso problem which can be solved by a classical Lasso solver \citep{lee2006efficient}.

\subsubsection{Dictionary optimization step}

Here we illustrate the estimation of $\{\mathbf{D}_{p}\}_{p=1}^{C}$ while fixing $\{\mathbf{a}_{n}\}_{n=1}^{N}$. It can be seen that (\ref{eq:2}) involves quadratic terms with respect to the dictionaries. The derivative of the objective function with respect to $\mathbf{D}_{p}$ is:

\begin{equation}
 \nabla_{\mathbf{D}_{p}} J= \nabla_{\mathbf{D}_{p}} J_{1}+ \nabla_{\mathbf{D}_{p}} J_{2}+ \gamma_{2} \nabla_{\mathbf{D}_{p}} J_{5} 
\label{eq:der}
 \end{equation}
with the involved terms defined below using the matrix derivation formula \citep{petersen2008matrix}. 
\begin{equation}
 \left\{
    \begin{array}{ll}
J_{1}=\displaystyle \sum_{n=1}^{N} \norme{\mathbf{x}_{n}-\mathbf{D} \mathbf{a}_{n}}_{2}^{2}= \sum_{n=1}^{N} \norme{ \tilde{\mathbf{x}}_{n}-\mathbf{D}_{p} \mathbf{a}_{np}}_{2}^{2} \\ 
& \\
\nabla_{\mathbf{D}_{p}} J_{1}= \displaystyle \sum_{n=1}^{N} -2 \mathbf{\tilde{x}}_{n} \mathbf{a}_{np}^{T}+2 \mathbf{D}_{p} \mathbf{a}_{np} \mathbf{a}_{np}^{T}
\end{array}
\right.
\label{eqq1}
\end{equation}
\vspace{0.2cm}
where  $\displaystyle \tilde{\mathbf{x}}_{n}=\mathbf{x}_{n}-\sum_{\substack {c=1 \\ c \neq p}}^{C} \mathbf{D}_{c} \mathbf{a}_{nc}$

\noindent For the second term of the derivative  $\nabla_{\mathbf{D}_{p}} J$ we can write

\begin{equation}
 \left\{
    \begin{array}{ll}
J_{2}=\displaystyle \sum_{n=1}^{N}  \mathbbm{1}_{y_{n} = p}  \norme{\mathbf{x}_{n}-\mathbf{D}_{p} \mathbf{a}_{np}}_{2}^{2} + \sum_{n=1}^{N} \sum_{c \neq p}   \mathbbm{1}_{y_{n}=c} \norme{\mathbf{x}_{n} - \mathbf{D}_{c} \mathbf{a}_{nc}}_{2}^{2}                \\
& \\
\nabla_{\mathbf{D}_{p}} J_{2}= \displaystyle \sum_{n=1}^{N}  \mathbbm{1}_{y_{n} = p} -2 \mathbf{x}_{n} \mathbf{a}_{np}^{T}+2 \mathbf{D}_{p} \mathbf{a}_{np} \mathbf{a}_{np}^{T} 
\end{array}
\right.
\end{equation}

\noindent Finally the expression of the last term is given by

\begin{equation}
 \left\{
    \begin{array}{ll}
J_{5}=\displaystyle \sum_{c \neq p} 2 \norme {\mathbf{D}_{p}^{T} \mathbf{D}_{c}}_{F}^{2} + \sum_{c \neq p} \sum_{\substack {c' \neq c \\ c' \neq p}} \norme {\mathbf{D}_{c}^{T} \mathbf{D}_{c'}}_{F}^{2} \\
& \\
\nabla_{\mathbf{D}_{p}} J_{5}= \displaystyle \sum_{c \neq p} 4 (\mathbf{D}_{c} \mathbf{D}_{c}^{T}) \mathbf{D}_{p}
\end{array}
\right.
\label{eqq2}
\end{equation}

\vspace{0.5cm}


Algorithm \ref {alg:1} summarizes the different steps of our optimization approach which is based on an alternating scheme: the first step consists of a signal sparse coding based on the Lasso algorithm. The second step is dictionary optimization based on proximal gradient descent approach. The proximal procedure is useful in order to handle the atom normalization constraint $\norme{\mathbf{d}_{ck}} \leq 1$ in the problem (\ref{eq:2}).

\begin{algorithm} [!h]
\begin{algorithmic} [1]
\STATE\textbf{ Initialization:}  $\mathbf {D}_{0}$, $t \leftarrow 1$, \mbox{initialize} $\eta_{0}$ and $\alpha$
\WHILE {$t \leq T$}
\STATE \mbox{Solve for} $\mathbf{A}_{t} \leftarrow \displaystyle \argmin_{\mathbf{A}} J(\mathbf{D}_{t-1},\mathbf{A})$ \mbox{using Lasso algorithm}\\
\STATE \mbox{Compute the gradient} $ \mathbf{G}_{\mathbf{D}_{t-1}} \leftarrow \nabla_{\mathbf{D}} J(\mathbf{D}_{t-1}, \mathbf{A}_{t})$ \mbox{based on equations (\ref{eq:der}) to (\ref{eqq2})}\\
\STATE $\eta \leftarrow \eta_{0}$ 
\REPEAT
\STATE $ \mathbf{D}_{\frac{t}{2}} \leftarrow \mathbf{D}_{t-1} - \eta \mathbf{G}_{\mathbf{D}_{t-1}}$ \\
\STATE $\mathbf{D}_{t} \leftarrow \mbox{Prox} \big(\mathbf{D}_{\frac{t}{2}}\big)$ \\

\begin{equation*}
 \mbox{with} \hspace {0.8cm} \mbox{Prox} \big(\mathbf{D}_{\frac{t}{2}}\big) :   \{\mathbf{d}_{k}\}_{k=1}^{K}= \left\{
    \begin{array}{l}
   \mathbf{d}_{k} \hspace{8mm} \mbox{if}   \hspace{4mm} \norme{\mathbf{d}_{k}}_{2} \leq 1 \\
        \\
       \frac{\mathbf{d}_{k}}{\norme{\mathbf{d}_{k}}_{2}} \hspace{4mm} \mbox{otherwise} 
    \end{array}\right.
    \end{equation*}
    \STATE $\eta \leftarrow \eta \times \alpha$
\UNTIL {$J(\mathbf{D}_{t},\mathbf{A}_{t})<J(\mathbf{D}_{t-1},\mathbf{A}_{t-1})$}
\STATE $t \leftarrow t+1$
\ENDWHILE
\end{algorithmic}
\caption{The optimization algorithm}
\label{alg:1}
\end{algorithm}

\FloatBarrier

\subsection{Classification}

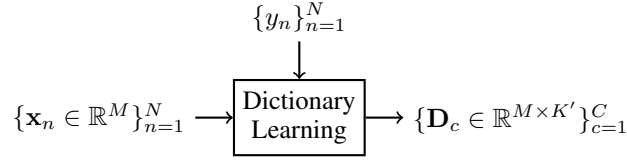
\begin{figure}[!h]
\centering
\begin{tikzpicture}  [thick,scale=1, every node/.style={scale=1}]
\tikzstyle{box} = [rectangle,draw,thick,align=center,minimum height=10mm];
\tikzstyle{arrow} = [->,thick];

\node[] (d) {$\{\mathbf{x}_{n} \in \mathbb{R}^{M}\}_{n=1}^{N}$};

\node[box,right=5mm of d.east,anchor=west] (tf) {Dictionary\\ Learning};

\node[right=5mm of tf.east,anchor=west] (dicolearn) {$\{\mathbf{D}_{c} \in \mathbb{R}^{M \times K'}\}_{c=1}^{C}$};

\node[above=5mm of tf.north,anchor=south] (target) {$\{y_{n}\}_{n=1}^{N}$};
\draw[arrow] (target)--(tf);

\draw[arrow] (d)--(tf);
\draw[arrow] (tf)--(dicolearn);

\end{tikzpicture}
\caption{Processing flow of dictionary learning on the training set.} \label{fig:feature_selection_contribution1}
\end{figure}

\begin{figure}[!h]
\centering
\begin{tikzpicture}  [thick,scale=1, every node/.style={scale=1}]
\tikzstyle{box} = [rectangle,draw,thick,align=center,minimum height=10mm];
\tikzstyle{arrow} = [->,thick];

\node[] (d) {$\{\mathbf{x}_{n} \in \mathbb{R}^{M}\}_{n=1}^{N}$};

\node[box,right=5mm of d.east,anchor=west] (dico) {Sparse Representation};

\node[above=5mm of dico.north,anchor=south] (dicolearn) {$\{\mathbf{D}_{c} \in \mathbb{R}^{M \times K'}\}_{c=1}^{C}$};

\node[box,right=20mm of dico.east,anchor=west] (svm) {SVM\\ Learning};

\node[above=5mm of svm.north,anchor=south] (target) {$\{y_{n}\}_{n=1}^{N}$};

\node[right=5mm of svm.east,anchor=west] (dddd) {$h$};

\draw[arrow] (d)--(dico);
\draw[arrow] (dicolearn)--(dico);
\draw[arrow] (dico)--(svm) node[above,pos=0.5] {$\{\mathbf{a}_{n}\}_{n=1}^{N}$};
\draw[arrow] (svm)--(dddd);
\draw[arrow] (target)--(svm);

\end{tikzpicture}
\caption{Processing flow of SVM training over the learned dictionary and training set.} \label{fig:feature_selection_contribution2}
\end{figure}
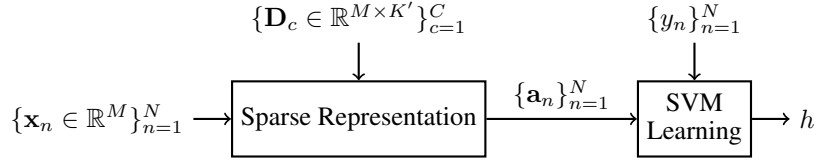

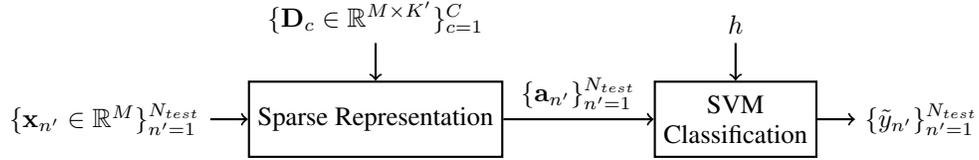
\begin{figure}[!h]
\centering
\begin{tikzpicture}  [thick,scale=1, every node/.style={scale=1}]

\tikzstyle{box} = [rectangle,draw,thick,align=center,minimum height=10mm];
\tikzstyle{arrow} = [->,thick];

\node[] (d) {$\{\mathbf{x}_{n'} \in \mathbb{R}^{M}\}_{n'=1}^{N_{test}}$};

\node[box,right=5mm of d.east,anchor=west] (dico) {Sparse Representation};

\node[above=5mm of dico.north,anchor=south] (dicolearn) {$\{\mathbf{D}_{c} \in \mathbb{R}^{M \times K'}\}_{c=1}^{C}$};

\node[box,right=20mm of dico.east,anchor=west] (svm) {SVM\\ Classification};

\node[above=5mm of svm.north,anchor=south] (svmlearn) {$h$};

\node[right=5mm of svm.east,anchor=west] (dddd) {$\{\tilde{y}_{n'}\}_{n'=1}^{N_{test}}$};

\draw[arrow] (d)--(dico);
\draw[arrow] (dicolearn)--(dico);
\draw[arrow] (dico)--(svm) node[above,pos=0.5] {$\{\mathbf{a}_{n'}\}_{n'=1}^{N_{test}}$};
\draw[arrow] (svm)--(dddd);
\draw[arrow] (svmlearn)--(svm);

\end{tikzpicture}
\caption{Processing flow of classification over testing set.} \label{fig:feature_selection_contribution3}
\end{figure}

Once the dictionaries are learned, they are used to encode both training and testing samples based on Lasso. The resulting coefficients are used to feed an SVM classifier. Figures \ref{fig:feature_selection_contribution1} to \ref{fig:feature_selection_contribution3} show the processing flow of dictionary learning based on the training data, coding both training and testing data over the learned dictionary respectively.

\FloatBarrier

Let define $\mathcal{H}$ a Hilbert space induced by  kernel $k(.,.)$. The decision function of a binary classification problem is given by  $h(\mathbf{a})=h_{0}(\mathbf{a})+b$ with $h_{0} \in \mathcal{H}$, $b \in \mathbb{R}$ and $\norme{h}_{\mathcal{H}}^{2}=\norme{h_{0}}_{\mathcal{H}}^{2}$ and is obtained as the solution of \citep{scholkopf2002learning}: 
\begin{equation}
 \left\{
    \begin{array}{ll}
       \displaystyle \min_{h_{0},b} \frac{1}{2} \norme{h}^{2}_{\mathcal{H}} + C_{svm} \sum_{n=1}^{N} \xi_{n} \\
       & \\
    \mbox{s.t}  \hspace{0.5 cm} {y}_{n} h(\mathbf{a}_{n})\geq 1-\xi_{n}, \hspace{0.3 cm} \xi_{n}  \geq 0 \hspace{0.2 cm} \forall n=1, \cdots,N\\
   \end{array}
\right.
\end{equation}
where $\Big\{ (\mathbf{a}_{n},y_{n}) \in \mathcal{A} \times \{ -1,+1\} \Big \}_{n=1}^{N} $ are the labelled training samples. $\xi_{n}$ and $C_{svm}$ represent slack variables and tuning parameter used to balance margin and training error. The solution is given by $\displaystyle h_0(\mathbf{a}) = \sum_{n=1}^{N} \alpha_n y_n k(\mathbf{a}_{n}, \mathbf{a})$ where parameters $\alpha_n$ are solution of the dual quadratic problem:

\begin{equation}
 \left\{
    \begin{array}{ll}
       \displaystyle \max_{\pmb {\alpha}}  \sum_{n=1}^{N} \alpha_{n} - \frac {1}{2} \sum_{n=1}^{N}  \sum_{n'=1}^{N} \alpha_{n} \alpha_{n'} y_{n} y_{n'}  k(\mathbf{a}_{n},\mathbf{a}_{n'})\\
       & \\
       \mbox{s.t}  \hspace{0.5 cm} \forall n \hspace{0.2 cm} 0 \leq \alpha_{n} \leq C_{svm}, \hspace{0.2 cm}  \displaystyle \sum_{n=1}^{N} \alpha_{n} y_{n}=0\\
   \end{array}
\right.
\end{equation}

To solve our $C$-class audio classification problem we employ one-against-all strategy.
It consists in constructing $C$ binary SVM, each one separates a class from all the rest.
The $c^{th}$ SVM solves the decision problem $h^{(c)}(\mathbf{a})=h_{0}^{(c)}(\mathbf{a})+b ^ {(c)}$ with data from class $c$ taken as positive samples and the remaining training samples as negatives. Note that in our case we have used a simple linear kernel as the non-linear aspect of the classification problem is taken into account in the dictionary learning. This is customary in supervised dictionary classification \citep{mairal2009,mairal2012task}.

\section{Experiments}
We conduct our experiments on two different audio signal classification problems, Computational Auditory Scene Recognition (CASR) and music chord recognition. For each problem, dictionary learning based on a initial time-frequency representation is compared to conventional hand-crafted features.

\subsection{Computational auditory scene recognition}
In this section we briefly review different approaches to tackle CASR problem as well as the evaluation of our proposed dictionary learning technique compared with conventional hand-crafted features on East Anglia (EA) and LITIS Rouen datasets.

\bigskip

Several categories of audio features have been employed in CASR systems. \citep{barchiesi2015acoustic} divided the features into 12 categories summarized in Table \ref{tab:summtfeaturescasr}. From the features organization in Table \ref{tab:summtfeaturescasr}, we can distinguish four main categories: low-level time/frequency, frequency band energy, learned features based on an time-frequency representation and speech-based. Among low-level features, we find easy and simple features to compute such as zero crossing \citep{eronen2006audio}. Frequency band energy feature are based on the computation of the energy at different frequency bands using Fourier transform  \citep{eronen2006audio} or filter banks such as Gammatone \citep{sawhney1997situational} and  Mel-scale filter banks \citep{clarkson1998auditory} which seek to mimic the response of the human auditory system. The goal of learning methods is to describe an acoustic signal as a linear combination of elementary functions that capture salient spectral components \citep {lee2013acoustic}. Beside the first three introduced feature categories, speech-based features and more particularly Mel-Frequency Cepstral Coefficients (MFCCs) represent the most prominent features that have been considered in the problem of audio scene recognition.

A considerable amount of works have applied MFCCs for CASR, \citep {aucouturier2007bag} used Gaussian Mixture Model (GMM) to estimate the distribution of MFCC coefficients. \citep {ma2006acoustic} combined MFCCs with Hidden Markov 

\begin{landscape}
\begin{table}[h]
  \centering
  \caption {Main audio feature categories for audio scene recognition \citep{barchiesi2015acoustic}.}
  \begin{tabular}{llll}
    \multicolumn{3}{c}{}  \\
        \toprule
         \toprule
Methods      & Approach  & Features  \\
    \midrule  
     \midrule  
  \citep{eronen2006audio}       &  Low-level time-based \& frequency-based    &  Zero crossing rate  \\
      \citep{malkin2005classifying}  &                                 & Spectral centroid   \\
                                   \hline
                                   \hline
\citep{eronen2006audio}   &    Frequency-band energy   & Magnitude or power spectrum  \\
                                     \hline
                                     \hline
 \citep{sawhney1997situational}    & Auditory filter banks  & Gammatone filters  \\
  \citep{clarkson1998auditory}        &                              & Mel-scale filter bank     \\
                                     \hline
                                      \hline
\citep{peltonen2002computational}   &        Cepstral     & Mel-frequency cepstral coefficients \\                                                      
                                         \hline
                                         \hline
\citep{nogueira2013sound}         &      Spatial      & Interaural time/level difference   \\
                                         \hline
                                         \hline
 \citep {krijnders2013tone}         &  Voicing   & Tone-fit features \\
                                         \hline
                                         \hline
  \citep{eronen2006audio}           &    Linear predictive model & Linear predictive coefficients  \\
                                         \hline
                                         \hline
                                         
  \citep{chu2009environmental}         &   Parametric approximation    & Convolution spectrogram and  \\
     \citep{patil2002multiresolution}      &                           &  Gabor filters \\
                                         \hline
                                         \hline                                        
                                         
   \citep {lee2013acoustic}            & Feature learning    & Learned features from MFCCs  \\
                                         \hline
                                         \hline
\citep{cauchi2011}         &  Matrix factorization     & Non-negative matrix factorization   \\
  \citep{benetos2012characterisation}      &   &  Probabilistic latent component \\
                                         \hline
                                         \hline
                                         
   \citep{rakotomamonjy2015histogram}             &  Image processing  &  HOG time-frequence representation  \\
                                         \hline
                                         \hline
  \citep {heittola2010audio}          &   Event detection   &  Analysis of events occurrence\\                                                                             
    \bottomrule
     \bottomrule
  \end{tabular}
  \label{tab:summtfeaturescasr}
\end{table}
\end{landscape}

\noindent Models (HMM). \citep{cauchi2011} used Non-Negative Matrix Factorization (NMF) with MFCC features. \citep{hu2012combining} employed MFCC features in a two-stage framework based on GMM and SVM. \citep{lee2013acoustic} used sparse restricted Boltzmann machine to capture relevant MFCC coefficients. \citep{geiger2013large} extracted a large set of features including MFCCs using a short sliding window approach. SVM is used to classify these short segments, and a majority voting scheme is employed for the whole sequence decision. \citep {roma2013recurrence} applied Recurrence Quantification Analysis (RQA) on the MFCCs for supplying some additional information on temporal dynamics of the signal.

Another trend is to extract discriminative features from time-frequency representations.  \citep{cotton2011spectral} applied NMF to extract time-frequency patches. \citep{benetos2012characterisation} instead of the NMF used temporally-constrained Shift-Invariant Probabilistic Latent Component Analysis (SIPLCA) to extract time-frequency patches from spectrogram. \citep{yu2008audio} proposed a method based on treating time-frequency representations of audio signals as image texture. In the same context, \citep{dennis2013image} introduced novel sound event image representation called Subband Power Distribution (SPD). The SPD captures the distribution of the sound's log-spectral power over time in each subband, such that it can be visualized as a two-dimensional image representation. Recently \citep{rakotomamonjy2015histogram} proposed to use Histogram of Oriented Gradient to extract information from time-frequency representations.

\subsubsection {Datasets}
We rely our experiments on two representative datasets which are described below.
\begin{itemize} 
\item East Anglia (EA): this dataset \footnote{\url{http://lemur.cmp.uea.ac.uk/Research/noise_db/}}
 provides environmental sounds \citep {ma2003context} coming from 10 different locations: \textit{bar, beach, bus, car, football match, launderette, lecture, office, rail station, street}. In each location a recording of 4-minutes at a frequency of 22.1 kHz has been collected. The 4-minutes recordings are splitted into 8 recordings of 30-seconds so that in total we have 10 locations (classes) and each class has 8 examples of 30-seconds. 
\end{itemize} 

\begin{itemize} 
\item Litis Rouen: this dataset \footnote{\url{https://sites.google.com/site/alainrakotomamonjy/home/audio-scene}} provides environmental sounds  \citep{rakotomamonjy2015histogram} recorded in 19 locations. Each location has different number of 30-seconds examples downsampled at 22.5 kHz. Table \ref{tab:liti} summarizes the content of the dataset.
\end{itemize}

\begin{table*} [h] \centering
 \caption{Summary of Litis Rouen audio scene dataset.}
\scalebox{1}{
\begin{tabularx}{13cm}{ Xc}
\hline
\hline
\textbf{Classes} &  \hspace{2mm} \textbf{\# examples} \\
\hline
\hline
plane  & 23  \\
busy street  & 143 \\
bus & 192\\
cafe       & 120    \\
car       &  243  \\
train station hall   & 269  \\
kid game hall    & 145   \\
market       &   276 \\
metro-paris       & 139  \\
metro-rouen       & 249   \\
billiard pool hall       & 155  \\
quite-street       &    90 \\
student hall       &  88  \\
restaurant      &   133 \\
pedestrian street       & 122   \\
shop      & 203    \\
train       &  164  \\
high-speed train  &  147 \\
tube station       &   125  \\
\hline
\hline
\end{tabularx}}
\label{tab:liti}
\end{table*}

\subsubsection {Competing features and protocols}
In the following we introduce the different features used in our experiments as well as the data partition and protocols.

\subsubsection*{Features}
Based on an initial time-frequency representation (spectrogram) computed on sliding windows of size $4096$ samples and hops of $32$ samples, we apply our class based dictionary learning method introduced in \ref{sec:ourcontribution}. In order to evaluate the efficiency of our proposed method, we compare its performance to the following conventional features:

\begin {itemize}

\item Spectrogram pooling: represents the temporal pooling of the spectrogram computed on sliding windows of size $4096$ samples and hops of $32$ samples.
\item Bag of MFCC: consists in calculating the MFCC features on windows of size $25$ ms with hops of $10$ ms. For each window, $13$ cepstra over $40$ bands are computed (lower and upper band are set to $1$ and $10$ kHz). The final feature vector is obtained by concatenating the average and standard deviation of the batch of $40$ windows with overlap of $20$ windows.
\item Bag of MFCC-D-DD: in addition of the average and standard deviation, the first-order and second-order differences of the MFCC over the windows are concatenated to the feature vector. 
\item Texture-based time-frequency representation: it consists on extracting features from time-frequency texture \citep{yu2008audio}.
\item Recurrent Quantification Analysis (RQA): aims to extract from MFCCs some additional information on temporal dynamics. For all MFCCs obtained over $40$ windows with overlap of $20$, $11$ RQA features have been computed \citep {roma2013recurrence}. Afterwards, MFCC features and RQA features are all averaged over time and MFCC averages, standard deviations as well as the RQA averages are concatenated to form the final feature vector.
\item HOG of time-frequency representation: applies HOG to time-frequency representations transformed to images. The time-frequency representations are calculated based on Constant-Q Transform (CQT). HOG is able to provide information about the occurrence of gradient orientations in the resulting images \citep{rakotomamonjy2015histogram}.
\end{itemize}
More details to extract these features can be found in  \citep{rakotomamonjy2015histogram}. Note that for classification, Support Vector Machine (SVM) with linear kernel is applied.

\subsubsection*{Protocols and parameters tuning}
For sake of comparison we have performed the same experiments using the same repartitions and protocols in \citep{rakotomamonjy2015histogram}. We have averaged the performances from $20$ different  splits of the initial data into training and test. The training set represents $80$ \% of data while the rest represents the test set. Our proposed dictionary learning technique requires the following parameters:

\begin {itemize}
\item $\lambda$, $\gamma_{1}$, $\gamma_{2}$ controlling respectively, the sparsity, the structure of sparse coefficients and pairwise orthogonality of learned dictionaries. The parameters are selected among $\{0.1, 0.2, 0.3\}$.
\item  $K^{'}$ the size of each dictionary $\mathbf{D}_{c}$. Its value is explored among $\{10, 20, 30\}$.
\end{itemize}

Beyond that we use a linear SVM classifier which its regularization parameter $C_{svm}$ is selected among $10$ values  logarithmically scaled between $0.001$ and $100$. All these parameters are tuned according to a validation scheme. Model selection is performed by resampling $5$ times the training set into learning and validation sets of equal size. The best parameters are considered as those maximizing the averaged performances on the validation sets. Note that K-SVD  \citep{aharon2006img} has been used to initialize the class based dictionaries and the parameters $T=200$,  $\alpha_{0}=0.5$ and $\eta=10^{-3}$ was applied for the optimization scheme (see Section \ref{sec:opt}).

\subsubsection{Results and analysis}

Table \ref{tab_ comp_casr} represents the performance (classification accuracy) comparison between different conventional features as reported in \citep{rakotomamonjy2015histogram} and our class based dictionary method on Rouen and EA datasets. Texture denotes the work of \citep{yu2008audio} while MFCC-D-DD denotes the MFCC with derivatives features. MFCC, MFCC-RQA, MFCC-900 and MFCC-RQA-900 denote, MFCC features, the MFCC with RQA with cut-off frequency of 10 kHz, the MFCC and the MFCC combined RQA with upper frequency set at 900 Hz respectively. Spectrogram pooling stands for the temporal pooling of the time-frequency spectrogram. HOG-full and HOG-marginalized represent the concatenation of histogram obtained from different cells resulting to very-high dimensionality feature vector and the concatenation of the averaged histograms over time and frequency respectively.

\begin{table*} [h] \centering
 \caption{Comparison of performances related to different feature representations on Rouen, EA  audio scene classification datasets. Bold values stand for best values on each dataset.}
\scalebox{0.8}{
\begin{tabularx}{18cm}{ X X l}
\hline
\hline
\textbf{Features} &  \hspace{2mm} \textbf{Rouen} &    \hspace {5mm} \textbf {EA} \\
\hline
\hline
Texture       &   \hspace{8mm} -  &   0.57 $\pm$ 0.13 \\
\hline
\hline
MFCC-D-DD  &  $0.66$ $\pm$ $0.02$ &   0.98 $\pm$ 0.04  \\
\hline
\hline
MFCC  &  0.67 $\pm$ 0.01  &   \textbf{1.00} $\pm$ \textbf{0.01}  \\
MFCC-900  &  0.60 $\pm$ 0.02 &   0.91 $\pm$ 0.07  \\
\hline
\hline
MFCC+RQA &  0.78 $\pm$ 0.01  &   0.95 $\pm$ 0.08 \\
MFCC+RQA-900 &  0.72 $\pm$ 0.02   &   0.93 $\pm$ 0.06  \\
\hline
\hline
HOG-full &   0.84 $\pm$ 0.01   &   0.99 $\pm$ 0.02   \\
HOG-marginalized &   \textbf{0.86} $\pm$ \textbf{0.01}   &   0.97 $\pm$ 0.06  \\
\hline
\hline
Spectrogram pooling &   0.85 $\pm$ 0.01   &   0.97 $\pm$ 0.04  \\
\hline
\hline
Dictionary learning &   0.71 $\pm$ 0.01   &   0.97 $\pm$ 0.04  \\
\hline
\hline
\end{tabularx}}
\label {tab_ comp_casr}
\end{table*}

\begin{figure}[!h]
\centering
\includegraphics [width= 15 cm] {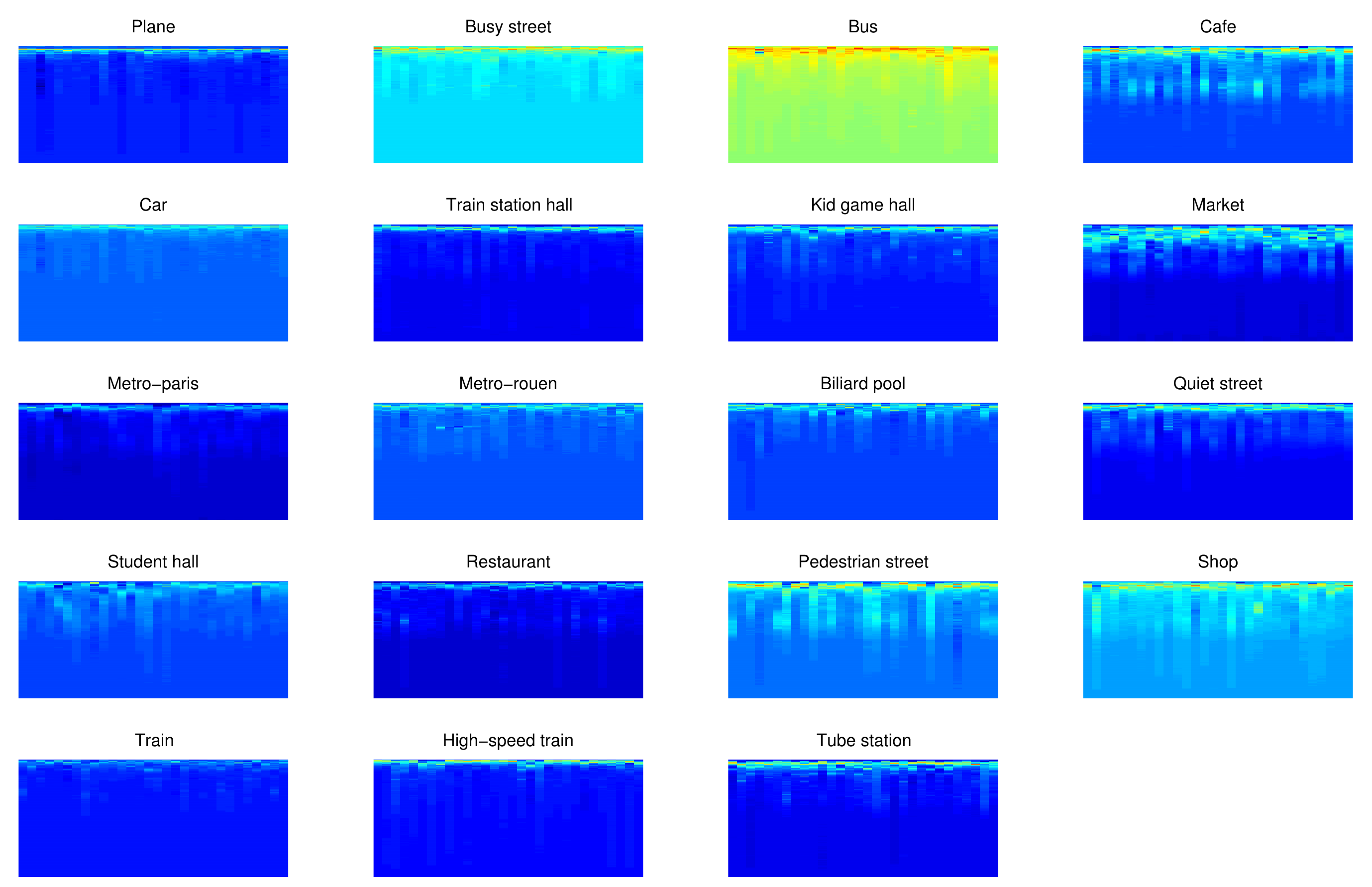}
\caption{Example of learned dictionaries per class on Rouen dataset. Rows correspond to learned dictionary atoms.}
\label{fig:rouendictionaries}
\end {figure}

\begin{figure}[!h]
\centering
\includegraphics [width= 9 cm] {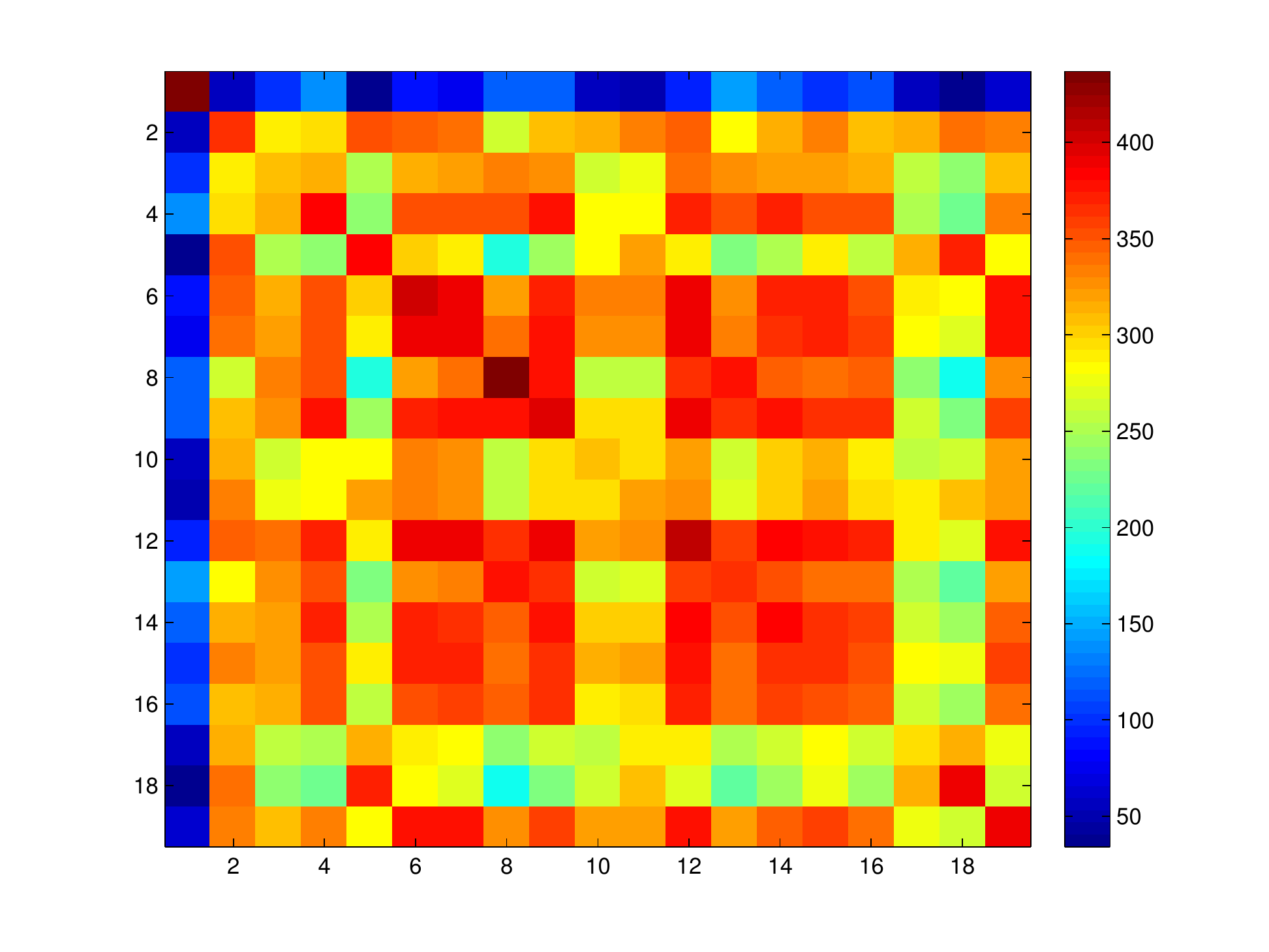}
\caption{Similarity between different learned dictionaries on Rouen dataset. X-axis and Y-axis stand for the class numbers organized in the same order in Table \ref{tab:liti}.}
\label{fig:rouensimilarity}
\end {figure}

It can be seen in Table \ref{tab_ comp_casr} that HOG-marginalized outperforms all competing features in Rouen dataset, it can be also seen that the temporal pooling of spectrogram is also giving good results and almost reach the ones obtained by HOG-marginalized. Surprisingly the temporal pooling of the spectrogram on all analysis windows helps to estimate the energy variation over time for a raw signal assumed to represent a single scene. Indeed it has been found that the use of the analysis windows improves the recognition performance. Moreover the small size of the windows helps to capture the stable characteristics of the signal \citep{pitchhistogram}. Note also that MFCC+RQA features are performing better than other MFCC based features, however the cut-off-frequency of $900$ Hz leads to a large loss in performance.

We can also notice that our proposed dictionary learning is giving very promising results and is outperforming texture and conventional speech recognition feature, MFCC and MFCC-D-DD features which have been widely used in the literature and have showed their ability to tackle the problems of audio scene recognition.

Figure \ref{fig:rouendictionaries} and Figure \ref{fig:rouensimilarity} show the learned dictionaries per  class on Rouen dataset and the pairwise similarity between them. The idea behind estimating the similarity between different learned dictionaries is to verify the initial goal to learn dissimilar dictionaries able to extract diverse information from classes for discrimination purpose. It can be seen that there is some similarity between some learned dictionaries which could influence the classification accuracy since these dictionaries tend to provide similar information for different classes. This may be related to the increasing number of classes that makes enforcing the pairwise dictionaries dissimilarity hardly feasible. 

In the East Anglia dataset, all features including our proposed dictionary learning perform well except texture, however we should note a slight advantage of MFCC.

\subsection {Music chord recognition}
The simplest definition of a chord is few musical notes played at the same time. In western music, each chord can be characterized by the:

\begin{itemize}
 
 \item  \textit {root or fundamental:}  the fundamental note on which the chord is built,
 \item   \textit{number of notes}
 \item \textit {type}: gives the interval scheme between notes. 
  
\end{itemize}

A music signal can be deemed composed of sequences of these different chords. Commonly, the duration of the chords in the sequence varies over time  rendering  their recognition difficult. Given a raw audio signal, chord recognition system attempts to automatically determine the sequence of chords describing the harmonic information.

To recognize chords most approaches rely on features crafted based on time-frequency representation of the raw signals, the most common and dominant features being chroma \citep{oudretemplate}. Pitch Class Profiles (PCP) or chroma vectors was introduced by \citep {chroma}. It is a 12-dimensional vectors representing the energy within an equal-tempered chromatic scale $ \{C , C^ {\#} , D ,\cdots, B \}$. The chroma has several variations, among them we can cite Harmonic Pitch Class Profiles (HPCPs) which is an extension of the Pitch Class Profiles (PCPs) by estimating the harmonics \citep {papadopoulos2008simultaneous,papadopoulos2007large} and Enhanced Pitch Class Profile (EPCP) which is calculated using the harmonic product spectrum \citep{lee2006automatic}. Chroma vectors were combined with different machine learning such as Hidden Markov and Support Vector Machine \citep{sheh2003chord,weller2009structured}.

\subsubsection {Dataset}

We will focus on third, triad and seventh chords which are respectively composed of 2, 3 and 4 notes.
When a note B has twice the frequency of a note A, the interval $[A\;B]$ forms an octave.
In tempered occidental music, the smallest subdivision of an octave is a semitone which corresponds to one twelfth of an octave, that is a multiplication by $\sqrt[12]{2}$ in term of frequency.
To be tertian, i.e a standard harmony, each interval between notes in a chord must be composed of 3 or 4 semitones.
These intervals are respectively called \textit{minor} and \textit{Major}.
Thus, for a given root, there is 2 possible thirds, 4 possible triads, and 8 possible sevenths.
Table \ref{tab:typeofchords} sum-up all the possible tertian third, triad and seventh chords.

\begin{table}[!ht]
\caption{Different kind of tertian chords, intervals are in semitones}
\label{tab:typeofchords}
\begin{tabular}{ccccc}
\hline
\hline
\# of notes & Common name or type & 1st interval &  2nd int. & 3rd int.\\
\hline
\hline
2 & Minor third & 3 & - & -\\
2 & Major third & 4 & - & -\\
\hline
3 & Diminished triad& 3 & 3 & -\\
3 & Minor triad& 3 & 4 & -\\
3 & Major triad& 4 & 3 & -\\
3 & Augmented triad& 4 & 4 & -\\
\hline
4 & Diminished seventh& 3 & 3 & 3\\
4 & Half-diminished seventh& 3 & 3 & 4\\
4 & Minor seventh& 3 & 4 & 3\\
4 & Minor major seventh& 3 & 4 & 4\\
4 & Dominant seventh& 4 & 3 & 3\\
4 & Major seventh& 4 & 3 & 4\\
4 & Augmented major seventh& 4 & 4 & 3\\
4 & Augmented augmented seventh & 4 & 4 & 4\\
\hline
\hline
\end{tabular}
\end{table}

The pursued goal in this work is to guess the type and not the fundamental of a chord  leading to 14 possible labels ($=2+4+8$). 
For this purpose, we have created a dataset which contains 2156 music chord samples of duration $2$-seconds at frequency $44100$~Hz with the 14 different classes.
Each class contains 154 samples from different instruments at different fundamentals.

\subsubsection {Competing features and protocols}

In the following we introduce the different features used in our experiments as well as the data partition and protocols.

\subsubsection*{Features}
Similar to the previous application we compute an initial time-frequency representation (spectrogram) on sliding windows of size 4096 samples and hops of 32 samples. Then we apply our dictionary learning method. The resulting sparse representations are used as inputs of an SVM. The following conventional features serve as competitors to our approach.

\begin {itemize}
\item Spectrogram pooling: represents the temporal pooling of the spectrogram as previously.
\item Interpolated power spectral density: music notes follow an exponential scale, however Power Spectral Density (PSD) is based on Fourier transform which follows a linear scale. To address this problem PSD (which lies on a linear scale) is sampled at specific frequencies corresponding to 96 notes leading to an exponential representation more suitable for chord recognition \citep{rida2014supervised}.
\item Chroma: it represents a $12$-dimensional vector, every component represents the spectral energy of a semi-tone within the chromatic scale. Chroma vector entries are calculated by summing the spectral density corresponding to frequencies belonging to the same chroma \citep{oudretemplate}.
\end{itemize}

\subsubsection*{Protocols and parameters tuning}

We have averaged the performances from different 10 splits of the initial data into training and test. The training set represents 2/3 of data. Model selection is performed by resampling $2$ times the training set into learning and validation set of equal size. The best parameters are considered as those maximizing the averaged performances on the validation sets. Note that the parameters are chosen from the same intervals used above in the computational auditory scene recognition problem.

\subsubsection{Results and analysis}

Table \ref{tab:tab_ comp_chord} represents the performance (classification accuracy) comparison of evaluated features on music chord dataset. It can be seen that our dictionary learning method outperforms all other features. 

\begin{table*} [h] \centering
 \caption{Comparison of performances related to different feature representations on music chord dataset based on linear SVM. Bold value stands for best performance.}
\scalebox{1}{
\begin{tabularx}{10cm}{ Xc}
\hline
\hline
\textbf{Features} &  \hspace{2mm} \textbf{Music chord} \\
\hline
\hline
Chroma  & 0.19 $\pm$ 0.01 \\
Interpolated PSD  & 0.15 $\pm$ 0.02 \\
Spectrogram pooling &  0.14 $\pm$ 0.01\\
Dictionary learning       &    \textbf{0.66 $\pm$ 0.01}  \\
\hline
\hline
\end{tabularx}}
\label {tab:tab_ comp_chord}
\end{table*}

Table \ref{tab:tab_ comp_chord_nonl} represents the performance (classification accuracy) comparison of evaluated features on music chord dataset based on the polynomial kernel. It can be seen the interpolated PSD outperforms chroma and spectrogram. It can be also noticed that the polynomial
kernel overcome the linear one in this particular task of chord recognition based on the conventional hand-crafted features.

\begin{table*} [h] \centering
 \caption{Comparison of performances related to different feature representations on music chord dataset based on polynomial kernel. Bold value stands for best performance.}
\scalebox{1}{
\begin{tabularx}{10cm}{ Xc}
\hline
\hline
\textbf{Features} &  \hspace{2mm} \textbf{Music chord} \\
\hline
\hline
Chroma  & 0.70 $\pm$ 0.01 \\
Interpolated PSD  & \textbf{0.74 $\pm$ 0.01} \\
Spectrogram pooling &  0.72 $\pm$ 0.01\\
\hline
\hline
\end{tabularx}}
\label {tab:tab_ comp_chord_nonl}
\end{table*}

\begin{figure}[!h]
\centering
\includegraphics [width= 15 cm] {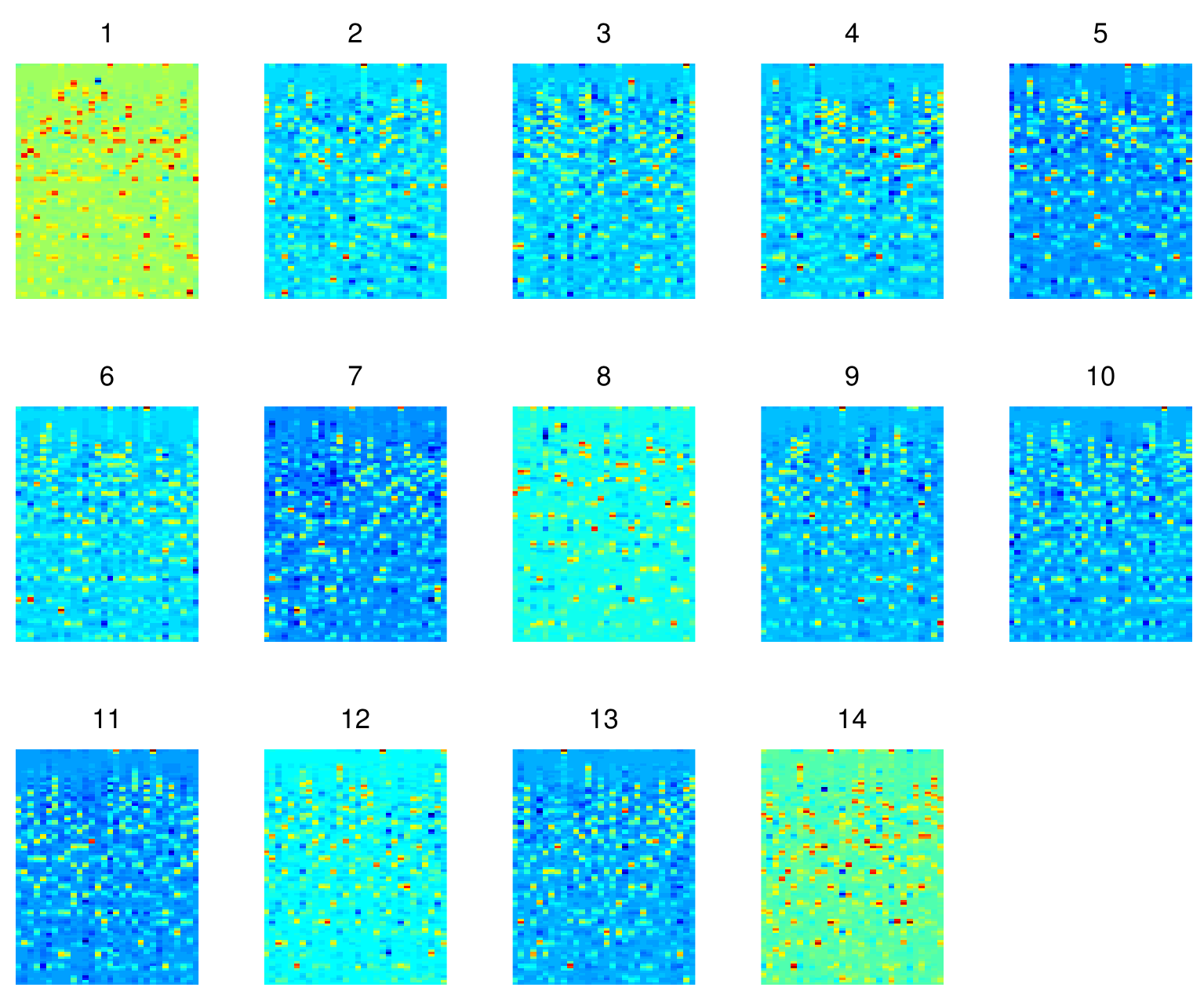}
\caption{Example of learned dictionaries per each class on music chord dataset.}
\label{fig:rouendictionariess}
\end {figure}

\begin{figure}[!h]
\centering
\includegraphics [width= 8 cm] {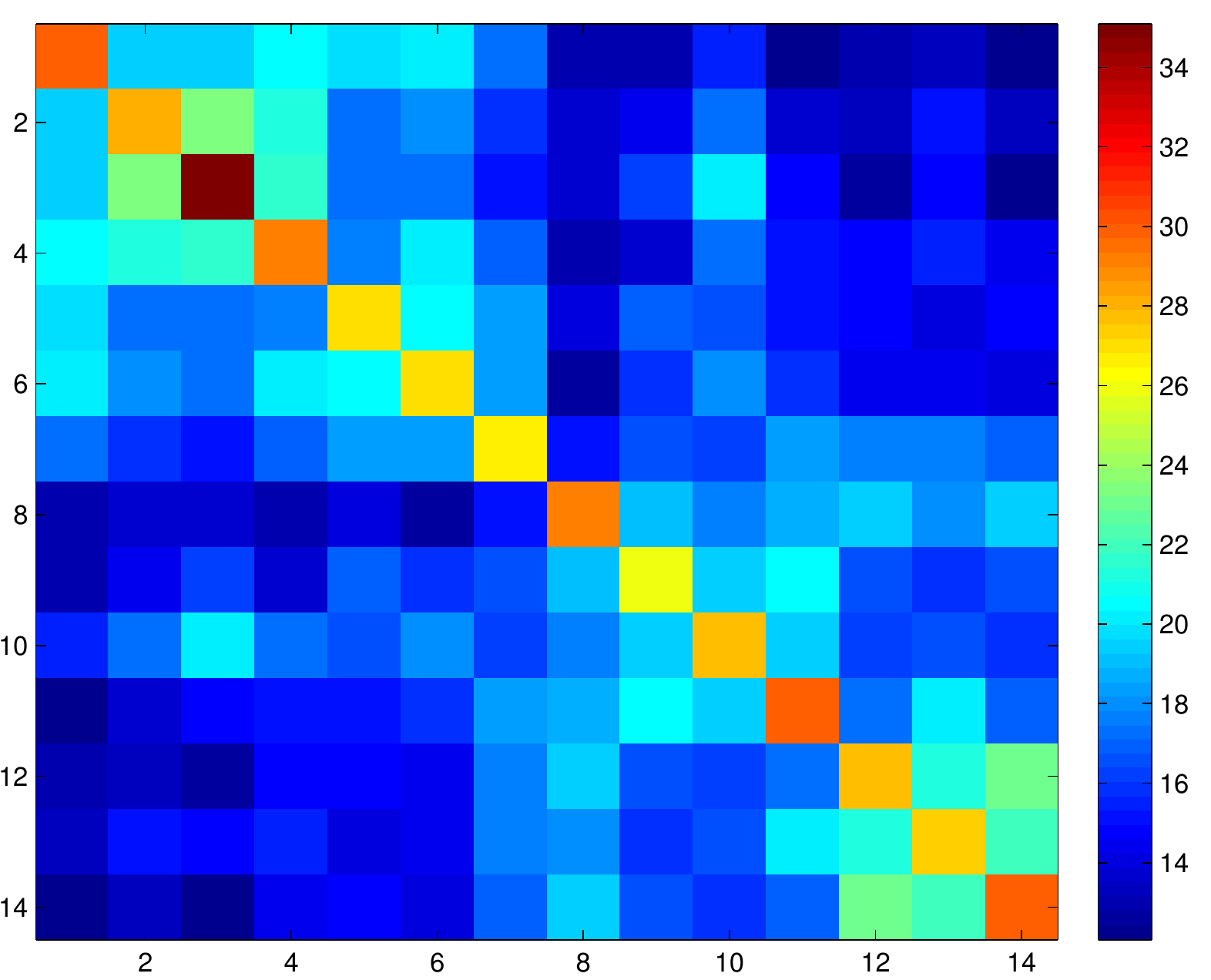}
\caption{Similarity between different learned dictionaries on music chord dataset. X-axis and Y-axis stand for the class numbers.}
\label{fig:rouensimilarityy}
\end {figure}

Figure \ref{fig:rouendictionariess} and Figure \ref{fig:rouensimilarityy} show the learned dictionaries and the pairwise similarity between them. Contrary to CASR Rouen dataset, it can be seen that the highest similarity between learned dictionaries is on the diagonal. This means that the  resulting dictionaries are different between them leading to extract diverse information per class. While chroma, interpolated PSD and spectrogram failed totally to reach good performances based on a linear SVM, our dictionary learning method could achieve very promising results.

Linear classification is a computationally efficient way to categorize test samples. It
consists in finding a linear separator between two classes. Linear classification has been the focus of much research in machine learning for decades and the resulting algorithms are well understood. However, many datasets cannot be separated linearly and require complex nonlinear classifiers which is the case of our music chord dataset. 

A popular solution to enjoy the benefits of linear classifiers is to embed the data into a high dimensional feature space, where a linear classifier eventually exists. The feature space mapping is chosen to be nonlinear in order to convert nonlinear relations to linear relations. This nonlinear classification framework is at the heart of the popular kernel-based methods \citep{shawe2004kernel}. Despite the popularity of kernel-based classification, its computational complexity at test time strongly depends on
the number of training samples \citep{burges1998tutorial}, which limits its applicability in large scale datasets.

An eventual alternative to kernel methods, is sparse coding which consists in finding a compact
representation of the data in an overcomplete learned dictionary which can be seen as a nonlinear feature representation mapping. This is confirmed by our experiments which clearly shows that our proposed dictionary learning method outperforms the other hand-crafted features. A success story of automatically learning useful features is represented by deep learning techniques \citep{bengio2013,lecun2015deep} which aim to learn several hierarchical layers, each layer can be seen as a kind of mapping operation to the one from dictionary learning.

\FloatBarrier
\section{Conclusion}

We have proposed a novel supervised dictionary learning method for audio signal recognition. The proposed method seek to minimize the intra-class variations, maximize the inter-class variations and promote the sparsity to control the complexity of the signal decomposition over the dictionary. This is done by learning a dictionary per class, minimizing the class based reconstruction error and promoting the pairwise orthogonality of the dictionaries. The learned dictionaries are supposed to provide diverse information per class. The resulting problem is non-convex and solved using a proximal gradient descent method.

Our proposed method was extensively tested on two different audio recognition applications: computational auditory scene recognition and music chord recognition. The obtained results were  compared to different conventional hand-crafted features.  While there is no universal hand-crafted feature representation able to successfully tackle different audio recognition problems, our proposed dictionary learning method combined with a simple linear classifier showed very promising results while dealing with the two diverse recognition problems.

Despite the simplicity and good performances of our approach, we could notice that the task to make the learned dictionaries as different as possible is hardly feasible when dealing with large number of classes. An example is human identity recognition based on gait where each individual is seen as a class. 

A possible alternative is to jointly learn the dictionary and classifier by incorporating a classification cost term. However, this will be leading to many parameters to tune, which makes the approach computationally expensive.


\end{document}